\documentclass[prb,showpacs,twocolumn,superscriptaddress]{revtex4-1}
\usepackage{amsmath,latexsym,bm,psfrag,color}
\usepackage[pdftex]{graphicx}
\usepackage{epstopdf}

\begin{document}

\title{Two-dimensional electron gas at the PbTiO$_3$/SrTiO$_3$ interface: 
an ab initio study}

\author{Binglun Yin}
\address{Department of Engineering Mechanics, Zhejiang University, 
310027 Hangzhou, China}
\address{CIC Nanogune, Tolosa Hiribidea 76, 20018 San Sebastian, Spain}

\author{P. Aguado-Puente}\email{p.aguado@nanogune.eu}
\address{CIC Nanogune, Tolosa Hiribidea 76, 20018 San Sebastian, Spain}
\address{Donostia International Physics Center DIPC, Manuel de
Lardizabal 4, 20018 San Sebastian, Spain}

\author{Shaoxing Qu}\email{squ@zju.edu.cn}
\address{Department of Engineering Mechanics, Zhejiang University, 
310027 Hangzhou, China}

\author{Emilio Artacho}
\address{CIC Nanogune, Tolosa Hiribidea 76, 20018 San Sebastian, Spain}
\address{Donostia International Physics Center DIPC, Manuel de
Lardizabal 4, 20018 San Sebastian, Spain}
\address{Theory of Condensed Matter, Cavendish Lab., 
University of Cambridge, JJ Thomson Ave, Cambridge CB3 0HE, UK}
\address{Basque Foundation for Science Ikerbasque, 48011 Bilbao, Spain}

\date{6 Aug 2015}

\begin{abstract}
  In the polar catastrophe scenario, polar discontinuity accounts for the 
driving force of the formation of a two-dimensional electron gas (2DEG) 
at the interface between polar and non-polar insulators.
  In this paper, we substitute the usual, non-ferroelectric, polar material 
with a ferroelectric thin film and use the ferroelectric polarization as the 
source for polar discontinuity. 
  We use ab initio simulations to systematically investigate the stability, 
formation and properties of the two-dimensional free-carrier gases formed 
in PbTiO$_3$/SrTiO$_3$ heterostructures under realistic mechanical and 
electrical boundary conditions.
  Above a critical thickness, the ferroelectric layers can be stabilized in 
the out-of-plane monodomain configuration due to the electrostatic 
screening provided by the free-carriers. 
  Our simulations also predict that the system can be switched between 
three stable configurations (polarization up, down or zero), allowing the 
non-volatile manipulation of the free charge density and sign at the interface.
  Furthermore, the link between ferroelectric polarization and free charge 
density demonstrated by our analysis constitutes compelling support for 
the polar catastrophe model that is used to rationalize the formation of 
2DEG at oxide interfaces.
\end{abstract}


\maketitle

\section{Introduction}

  Today, developments in deposition techniques allow researchers to routinely 
grow atomically sharp interfaces between transition metal oxides, 
boosting research on emergent phenomena at complex oxide 
interfaces.\cite{1,2} 
  Among these phenomena, one of the most striking examples is 
the formation of two-dimensional electron gas (2DEG) at the interface 
of two band insulators, LaAlO$_3$ (LAO) and SrTiO$_3$ (STO).\cite{3} 
  During the past decade, considerable research efforts have been 
devoted to explore the LAO/STO interface and similar systems, showing 
that at oxide interfaces 2DEG can exhibit enhanced capacitance,\cite{4} 
magnetism,\cite{5} superconductivity,\cite{6,7} or combinations of these 
properties.\cite{8,9} 
  The discovery of 2DEG and the possibility of its manipulation and 
coupling with other functional properties typical of oxide perovskites have 
fueled intense research activity aiming at exploiting its potential 
functionalities.
  Indeed, several potential practical applications based on 2DEG have 
been proposed, including field effect devices,\cite{10,11,12,13,14} 
sensors,\cite{15} and solar cells.\cite{16,17}

  Despite all these efforts, the driving force for the accumulation of the
free charge at these interfaces and the origin of charge itself are still 
controversial issues. 
  In the polar catastrophe scenario,18 a widely acknowledged model, 
the driving force for the formation of 2DEG is the mismatch of formal 
polarizations between the constituent materials.\cite{19}
  Such discontinuity has a huge electrostatic energy cost, and thus 
triggers a series of screening mechanisms that yield the accumulation 
of free charge at the interface.
  Among the various mechanisms that have been proposed, one of the 
most notables is the \emph{electronic reconstruction}.
  According to the conventional meaning of this term in the context of 
polar interfaces (see Ref. 20 for example), in pristine structures (with no 
defects, or surface adsorbates), this mechanism consists of a charge 
transfer between the valence and conduction bands of the system that 
results in an accumulation of free carriers at the interface.
  In practice, this mechanism might be accompanied by alternative 
sources of free charge, such as defects or surface redox 
processes.\cite{21,22,23}
  Nevertheless, regardless of the source of the free charge, the properties 
of the 2DEG can in principle be tuned by manipulating its driving force, 
i.e., the polar discontinuity.

  The polar discontinuity can be tuned in different ways. 
  For polar centrosymmetric materials like LAO, the polar discontinuity can 
be effectively changed by choosing different crystalline orientations for 
the interface,\cite{24} or diluting the polar material.\cite{25}
  After growth, the density of charge carriers at the interface can be tuned 
with an external electric field.\cite{10,11,12,13,14} 
  Another interesting possibility is to couple the 2DEG with a ferroelectric 
material, since the spontaneous polarization of the ferroelectric might allow 
the non-volatile manipulation of the electrostatic boundary conditions of the 
system.
  This approach was demonstrated in Ref. 26, where the authors observed a 
non-volatile metal-insulator transition in the LAO/STO interface after switching 
the spontaneous polarization of a ferroelectric layer deposited over the LAO.

  A more radical alternative is directly to replace the polar material with a 
ferroelectric layer, where the ferroelectric polarization could constitute the 
source for the polar discontinuity, instead of the ``built-inÓ polarization of a 
polar material such as LAO. 
  Such a ferroelectric system would possess some advantageous properties. 
  First, unlike in the case of LAO, the magnitude of the polar mismatch in 
ferroelectric interfaces can be increased or decreased, and even the sign 
of the discontinuity might be switchable with the polarization.
  Second, ferroelectric thin films are typically grown at a temperature higher 
than the Curie one, meaning that for usual materials (BaTiO3, PbTiO3, 
PbZr$_x$Ti$_{1-x}$O$_3$, etc.) there is no polar discontinuity during growth and therefore these systems 
should be less prone to the formation of charged defects.
  Of course this scenario also presents some problems.
  The switchable nature of ferroelectric polarization allows the system to find 
alternative routes to avoid or minimize the polar catastrophe, like forming 
domains or stabilizing in a paraelectric phase. 

  The possibility of a 2DEG at ferroelectric interfaces has been recently 
explored using first-principles calculations and phenomenological models.
  Simulations of symmetric KNbO3/ATiO3 (A=Sr, Ba, Pb) superlattices 
suggested that indeed the properties of 2DEG formed at ferroelectric 
interfaces can be modulated with polarization.\cite{27}
  Nevertheless the KNbO3 layers used in that paper were not stoichiometric 
and thus the superlattices were metallic by construction (intrinsic doping).
  Therefore, no conclusion about the stability of the 2DEG itself could be 
extracted from those results. 

  Very recently, two independent papers have provided strong arguments 
supporting the possibility of a 2DEG spontaneously forming to screen 
the depolarizing field in ferroelectric thin films and its subsequent 
manipulation through its coupling with the polarization.
  Ab initio simulations of BaTiO$_3$/STO/BaTiO$_3$ slabs have revealed 
that 10-unit-cell-thick BaTiO$_3$ layers can sustain a monodomain 
polarization with free carriers appearing at the surface and interface.\cite{28}
  However, this monodomain state was only found in one combination 
of polarization direction (pointing towards the interface) and surface 
termination (TiO$_2$-terminated), while all other configurations turned 
out to be paraelectric and insulating after ionic relaxation.
  In another independent study, a Landau-based model was used to 
discuss the stabilization of a monodomain ferroelectric phase thanks to 
the screening provided by a 2DEG spontaneously formed after an electronic
reconstruction.\cite{29}
  As in the case of the LAO/STO interface, the 2DEG (and consequently the 
monodomain ferroelectricity) only becomes stable above a critical thickness 
of the ferroelectric layer, which depends on the energetics associated with 
the mechanism that provides the free charge.
  The model even suggested that in the appropriate conditions the 
monodomain polarization screened by a 2DEG might be more stable than 
the polydomain one.
   Indeed, in experiments, monodomain polarization is routinely observed 
 in ferroelectric thin films on insulating substrates.\cite{30,31,32,33,34,35}
   Such a configuration, which can only be stable if the free charge accumulates at 
the interfaces, together with the two previously mentioned theoretical papers, 
provide a strong motivation to further investigate and characterize these systems.

  Here we perform first-principles simulations systematically to analyze 
the formation of two-dimensional (2D) electron and hole gases in a prototypical ferroelectric 
interface, PbTiO3/SrTiO3 (PTO/STO), under realistic mechanical and electrical 
boundary conditions.
  We investigate the transition with thickness from paraelectric to a tri-stable 
regime in which two polar (and metallic) and one non-polar (and insulating) 
configurations are accessible. 
  The properties of the stable monodomain structure and the 2D 
free-carrier gases are investigated, providing a detailed analysis of the free 
charge distribution. 

\section{Methodology}

  We carried out simulations of PTO/STO heterostructures using the 
formalism of the density functional theory as implemented in the 
SIESTA method.\cite{36}
  Exchange and correlation were treated within the local density 
approximation.
  Core electrons were replaced by ab initio norm-conserving, fully 
separable,\cite{37} Troullier-Martin pseudopotentials.\cite{38}
  For Pb atoms, the scalar relativistic pseudopotential was generated 
using the reference configuration 6$s$2, 6$p$2, 5$d$10, 5$f$0 with 
cut-off radii of 2.0, 2.3, 2.0, 1.5 atomic units, respectively.
  In SIESTA the one-electron eigenstates are expanded in a set of 
numerical atomic orbitals.
  The Pb basis set included a single-$\zeta$ basis set for the semicore 
5$d$ orbitals, a double-$\zeta$ for the valence 6$s$ and 6$p$ orbitals, 
and a single-$\zeta$ for two extra 6$d$ and 5$f$ polarization functions.
  The details about the pseudopotential and basis set of the rest of the 
atoms can be found elsewhere.\cite{39}
  A Fermi-Dirac distribution with a temperature of 100 K (~8.6 meV) was 
used to smear the occupancy of the one-particle electronic eigenstates.
  Reciprocal space integrations were performed on a Monkhorst-Pack\cite{40,41} 
$k$-point mesh equivalent to $6\times 6\times 6$ in a five-atom 
perovskite unit cell, while for real space integrations, a uniform grid 
with an equivalent plane-wave cutoff of 600 Ry was used.

  In this paper, the in-plane lattice constant in all calculations was fixed 
to that of the fully relaxed cubic STO in bulk, i.e., a = 3.874 \AA, to implicitly 
simulate the mechanical constraint imposed by the substrate. In constrained 
bulk PTO, the out-of-plane lattice constant is found to be c = 4.03 \AA, with 
a spontaneous polarization $P_S = 0.771$ C/m$^2$.
  Note that, $P_S$ here is obtained using the corrected Born effective 
charge method,\cite{42} which is the first-order approximation of the total 
polarization (both ionic and electronic contributions).
  The polarization obtained with this method is a good approximation 
to the exact value provided by the Berry phase formalism (0.776 C/m$^2$).

  For the PTO/STO heterostructures, the slab geometry with a generic 
formula SrO-(TiO$_2$-SrO)$_6$-(TiO$_2$-PbO)$_m$-vacuum was used, 
where $m$ is the thickness of the PTO layer and the vacuum thickness was 
set to approximately 50 \AA.
  Here, the substrate consisting of 6 unit cells of STO was found to be thick 
enough since the calculation with a thicker one (12 unit cells) yielded 
similar results (see Appendix A for further details).
  The dipole correction was used in all heterostructure calculations to 
avoid a spurious electric field due to the periodic boundary conditions.
  For the geometry optimization, the bottom three monolayers (7 atoms) of STO 
were fixed to the bulk structure to mimic the presence of a semi-infinite 
substrate.
  The out-of-plane atomic coordinates of the rest of the atoms were relaxed 
until the maximum force was less than 0.025 eV/\AA.

\section{Results}

\subsection{Frozen-phonon calculation}

  In order to start the geometry optimizations with reasonable structures, 
we followed the rationale of the phenomenological model presented in 
Ref.~\onlinecite{29} to predict the critical thickness $m_0$ for the stable 
ferroelectric configuration.
  According to Ref.~\onlinecite{29}, the free energy per unit area of a 
ferroelectric thin film in open boundary conditions can be expressed as
\begin{align}
G = &{ d \over 2 \epsilon_0 \chi_{\eta} } \left ( {P_{\eta}^4 \over 4 P_S^2}
- { P_{\eta}^2 \over 2} \right ) \\ &+ {d \over 2 \epsilon_0 \epsilon_{\infty} }
\left ( \sigma - P_{\eta} \right )^2 + \Delta \sigma + {\sigma^2 \over 2g}
\nonumber
\end{align}
  Here, $d$ corresponds to the thickness of the ferroelectric layer 
(defined as $d = mc$), $P_{\eta}$ and $\chi_{\eta}$ are the contributions 
of the ferroelectric soft mode to the polarization and electric susceptibility,
respectively, $P_S$ is the spontaneous polarization in bulk, $\epsilon_{\infty}$ 
is the background contribution of the relative permittivity,\cite{44,45}
$\epsilon_0$ is the vacuum permittivity, $\sigma$ is the area density of 
free carriers, $\Delta$ is the effective band gap (that should take into 
account the band alignment across the interface) and $g$ is the reduced 
density of states.\cite{22} 
  In Eq. (1) positive values of both $P_{\eta}$ and $\sigma$ are always 
assumed.
  Taking into account the electrostatic boundary conditions of the system, 
the total polarization of the ferroelectric layer, $P$, can be calculated as
\begin{equation}
P = { P_{\eta}+ ( \epsilon_{\infty} - 1 ) \sigma \over \epsilon_{\infty} }
\end{equation}
  Eq. (1) describes the energy balance between the tendency of a ferroelectric 
to develop an electric polarization [first term in the right-hand side of Eq. (1)], 
hindered by the interaction with a depolarizing field (second term of the 
equation), and the energy cost of creating free charge (which might be provided 
by different sources) that could partially screen the depolarizing field.
  The cost of forming the 2DEG is accounted for by the last two terms of Eq. (1) 
and comprises the energy required to promote electrons across the band 
gap of the system (taking into account the band alignment and other interfacial 
effects in the case of heterostructures) and the energy associated with the filling 
of the bands.
  A more detailed explanation about the original model can be found in 
Ref.~\onlinecite{29}. 

\begin{figure}
\begin{center}
\includegraphics[width=\columnwidth]{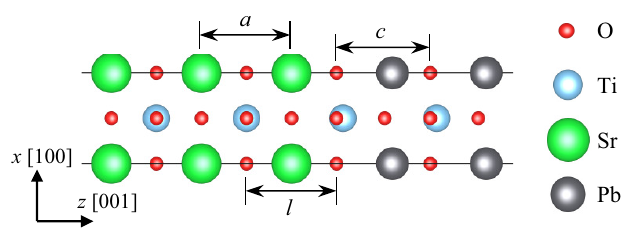}
\caption{\label{fig1} Schematic illustration of the interfacial atomic structure 
used in frozen-phonon calculations.
  In this paper, the distance l between the two oxygen atoms at the interface 
is set to $a$, the lattice constant of bulk STO.} 
\end{center}
\end{figure}

  According to the model, even in the absence of extrinsic screening, a 
metastable monodomain ferroelectric state might appear when the thickness
of the ferroelectric layer $d$ is larger than a critical value.
  The monodomain configuration would be accompanied by the formation 
of a 2D free-charge gas at the interfaces and/or surfaces of the 
ferroelectric material.
  Here we explore the gradual emergence of the local energy minimum 
utilizing a frozen-phonon method within the ab initio formalism.
  The details of the atomic structure used in our frozen-phonon calculation 
are depicted in Fig. 1.
  We construct the heterostructure by stacking m bulk unit cells of the 
ferroelectric material on top of 6 bulk unit cells of STO. 
  In doing so there is an inevitable arbitrariness in the choice of the interlayer 
distance at the interface.
  The influence of the interface construction is, according to the model, 
mostly through the band alignment and, therefore, the effective band gap 
of the system.\cite{29}
  However, since the phenomenon is primarily governed by electrostatics, 
any reasonably realistic choice for the interface structure should yield similar 
qualitative results.
  We confirmed that, indeed, different values of the interfacial distances, $l$ 
(see Fig. 1), produced similar estimations for the critical thickness $m_0$.
   In this paper the interfacial distance $l$ is set to be $a$, the lattice constant 
of bulk STO.
  Different values of the polarization in PTO layers are obtained by scaling 
the soft-mode distortion while keeping the lattice constants unchanged 
($c_{STO} = a = 3.874$ \AA, $c_{PTO} = c = 4.03$ \AA).
  Throughout this paper, a polarization along the [001] axis pointing towards 
(away from) the STO substrate is denoted as downward (upward) polarization, 
has a negative (positive) value and is labeled as $P_{\downarrow}$ 
($P_{\uparrow}$).
  The calculation of the total energy per unit area $G$ of the heterostructure 
as a function of polarization and PTO thickness results in the points shown 
in Fig. 2, where the gradual development of the``triple-wellÓ profile with 
increasing thickness is clearly reproduced, indicating a critical thickness 
for the stable ferroelectricity of $m_0 \sim 14$ for both polarization orientations.

\begin{figure}
\begin{center}
\includegraphics[width=\columnwidth]{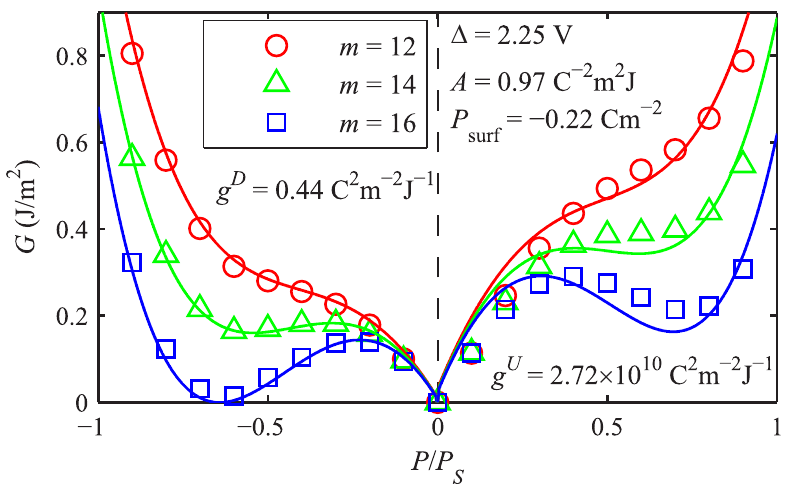}
\caption{\label{fig2} Free energy per unit area $G$ as a function of PTO 
polarization for different thickness $m$, as calculated using the 
frozen-phonon method. The curves are the least-squares fitting of 
Eq. (3), and the obtained fitting parameters are also shown.} 
\end{center}
\end{figure}

  The energy curves obtained with the frozen-phonon methods can be 
compared with the predictions of the model described in Ref.~\onlinecite{29}. 
  As shown in Fig. 2, the frozen-phonon results display an asymmetry with 
respect to the polarization orientation.
  The model in Ref.~\onlinecite{29} takes the same parameters for up and 
down polarization, but in general real interfaces do not show that symmetry. 
  In this case, for instance, the PTO film has vacuum on one side and a STO 
substrate on the other.
  This asymmetry might affect the shape of the $G(P)$ curves in different ways. 
  $(i)$ Free surfaces of these materials tend to develop a surface dipole that 
modifies the polarization near the surface relative to the bulk layers in 
relaxed slabs,\cite{43} slightly but noticeably favoring one polarization 
orientation over the other. 
  $(ii)$ In addition to this, the interface dipole, ultimately responsible for 
the band alignment at the interface, is also polarization-dependent, affecting 
the ``effective band gap" entering in the model equations and altering the 
competition between monodomain and paraelectric configurations.
  Here, to account for the asymmetry of the structure and the fact that in the 
frozen-phonon calculations the polarization is frozen to be homogeneous 
throughout the ferroelectric, we extend the original model\cite{29} by 
adding new terms that phenomenologically describe the tendency to 
develop a surface dipole.
  To first order, such effect can be described by a parabola centered at the 
polarization value corresponding to the surface dipole, $P_{\rm surf}$, 
and scaling with the area (not the volume) of the system.
  After incorporating this correction, the free energy per unit area, $G$, 
can be expressed as
\begin{align}
G = &{ d \over 2 \epsilon_0 \chi_{\eta} } \left ( {P_{\eta}^4 \over 4 P_S^2}
- { P_{\eta}^2 \over 2} \right ) + {d \over 2 \epsilon_0 \epsilon_{\infty} } \nonumber
\left ( \sigma - P_{\eta} \right )^2 \\ + & \; \Delta \sigma + {\sigma^2 \over 2g}
+ A \left ( P_{\eta} - P_{\rm surf} \right )^2 - A P_{\rm surf}^2 
\end{align}
where $A$ is the ``stiffness" of the surface dipole.
  The last term of $G$ is introduced in order to set to zero the energy 
reference at $P_{\eta} = 0$.

  With the version of the model described by Eq. (3) we performed a 
least-squares fitting of the curves obtained with the frozen-phonon 
approximation.
  Since the analysis of the electronic structure of fully relaxed interfaces 
(discussed below) reveals that the band alignment of this system is quite 
insensitive to polarization orientation, all six curves (for both polarization 
orientations and different thicknesses) were fitted for the same values of 
$\Delta$, $A$, and $P_{\rm surf}$.
  The reduced density of states ($g^D$ and $g^U$ for $P_{\downarrow}$ 
and $P_{\uparrow}$, respectively) was allowed to be different for each of 
the two polarization directions.
  The rest of the parameters in Eq. (3) were fixed to the bulk PTO values as 
obtained from independent ab initio calculations, i.e., $P_S = 0.771$ C/m$^2$, 
$\chi_{\eta} = 26$, $\epsilon_{\infty} = 7$.
  The least-squares fitting produced the solid curves and parameters shown 
in Fig. 2, demonstrating an excellent agreement between the frozen-phonon 
results and the model.
  Also, the obtained fitting parameters are all comparable to the bulk values, 
falling into a reasonable range for a realistic interface.
  The unrealistically large value of $g^U$ produced by the fit is attributed 
to the dependence of Eq. (3) on the inverse of the reduced density of states.
  With increasing $g$, the energy term associated with $g$ goes to zero very 
rapidly, meaning that if $g$ is relatively large, this energy term is negligible 
as compared with the rest of the contributions.
  In fact, for all practical purposes, a density of states larger than 
$\sim 2$ C$^2$m$^{-2}$J$^{-1}$ is indistinguishable from the limit 
of $g \rightarrow \infty$. 
  Notably, all the structures with non-zero polarization show metallic 
behavior, indicating that the electronic reconstruction only disappears 
within a tiny region around $P = 0$, a result that is also consistent with the 
model presented in Ref.~\onlinecite{29}.

\subsection{Polarization profile}

\begin{figure}
\begin{center}
\includegraphics[width=\columnwidth]{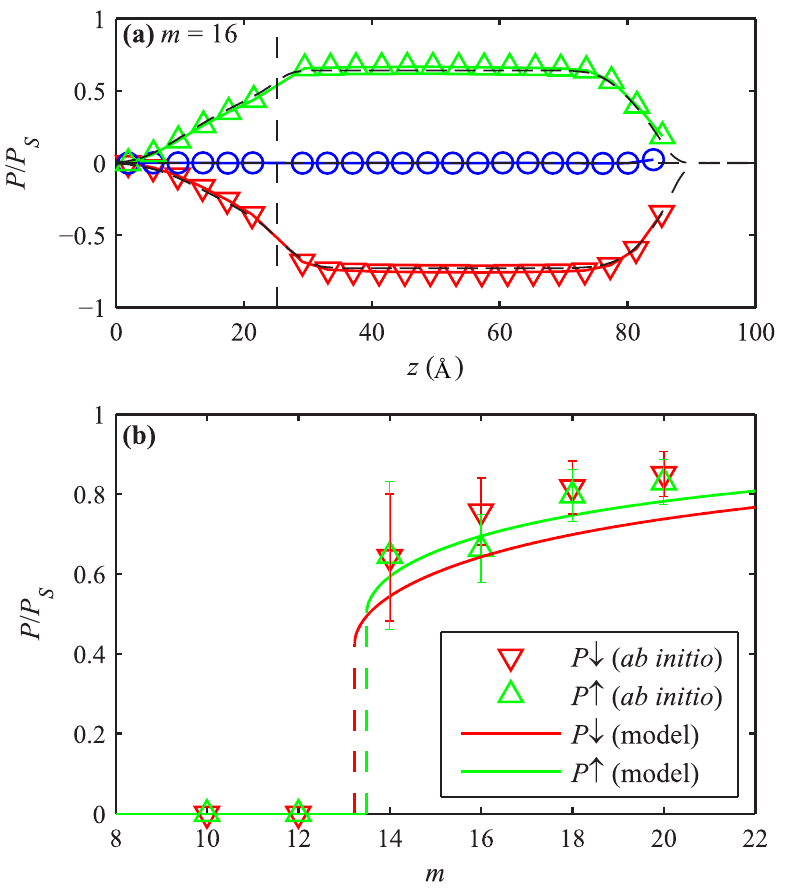}
\caption{\label{fig3} 
(a) Polarization as a function of $z$ for the PTO/STO heterostructures 
with $m = 16$ after ionic relaxation. It is calculated by computing the 
dipole at each Ti-centered unit cell using the corrected Born effective 
charge method.\cite{42} 
  The system is stable in either a ferroelectric state, with downward 
(red) or upward (green) polarization, or in a paraelectric one (blue).
  The black dashed curves indicate the electric field displacement as 
obtained from the integration of free carrier density.
  The vertical dashed line represents the position of the interfacial Ti 
atom in the paraelectric state. 
  (b) The magnitude of the stable ferroelectric polarization is plotted as 
a function of thickness. 
  Symbols correspond to fully relaxed structures.
  The continuous curves are obtained using the model described 
by Eq. (3) and the parameters from the fittings of Fig. 2, and not by 
fitting the polarization values of the relaxed structures.
  The error bars represent an estimation of the possible polarization 
fluctuations during ab initio calculations due to the shallowness of the 
ferroelectric energy minimum (details of their derivation are given in 
Appendix B).} 
\end{center}
\end{figure}

  To confirm the stability of the polar configurations, we performed ionic 
relaxations for the heterostructures with m ranging from 10 to 20 unit cells 
for both $P_{\downarrow}$ and $P_{\uparrow}$, as well as those with 
paraelectric PTO layers.
  The relaxed polarization profiles with $m = 16$ are shown in Fig. 3(a), 
and the rest are analogous.
  As shown in Fig. 3(a), for $m = 16$ the two polar configurations and 
the non-polar one are all stable and the polarization in the ferroelectric 
phase is uniform inside PTO, with values of $-0.76 P_S$ and $0.66 P_S$ 
for downward and upward polarizations, respectively.
  
  The evolution of the stable polarization with thickness is depicted in 
Fig. 3(b) alongside two curves obtained using the model for the two 
polarization orientations.
  These two curves were calculated minimizing Eq. (3) with respect to 
polarization and free charge in order to obtain their equilibrium values 
as a function of thickness.
  Then, the fitting parameters obtained from the frozen-phonon calculations, 
shown in Fig. 2, were used to obtain the solid curves in Fig. 3(b).
  Figure 3(b) evidences again the good agreement between model and 
first-principles calculations.
  The curves obtained from the model slightly underestimate the polarization 
values obtained in the relaxed structures.
  The reason for this is probably the fact that in the model and the 
frozen-phonon calculation, the polarization is assumed to be homogeneous 
throughout the ferroelectric layer.
  As a result, the pinning of the surface dipole restrains the bulk-like region 
of the ferroelectric film from developing a larger value.
  In addition to this, in the relaxed structures, atomic distortions in the 
STO contribute to the screening of the free charge injected in the substrate, 
increasing the density of carriers that the interface can host and therefore 
enabling the PTO to develop a larger polarization.
  All these effects combined, however, give rise to a relatively small deviation 
from the obtained ab initio results of the otherwise quite adequate model.
 
  Polarization values from ab initio relaxations are also susceptible to some 
fluctuations due to the shallowness of the energy minimum corresponding 
to the ferroelectric phase near the transition thickness.
  At the transition, the ferroelectric energy minimum becomes a saddle point, 
and the coordinate relaxation becomes an ill-defined problem.
  Away from the transition, convergence is possible, but polarization fluctuations 
occur for any finite threshold in the forces.
  In Appendix B, we derive an estimation of these fluctuations, which are 
represented as error bars in Fig. 3(b).
 The magnitude of these fluctuations is sizable near the transition but decays 
 rapidly as the thickness increases.
  
   The polar configurations in Fig. 3 directly demonstrate that in realistic 
boundary conditions, ferroelectric monodomain polarization can exist without 
any extrinsic screening mechanisms or polydomain formation.
  Furthermore, above some critical thickness $m_0$, the monodomain 
polarization can be stable in both downward and upward configurations, 
opening the door to a possible switching between 2DEG and two-dimensional 
hole gas (2DHG) at the ferroelectric interface.
   It is worth pointing out that, even if for this system the critical thickness is 
found to be the same for both polarization orientations, this is not necessarily 
true in general for asymmetric heterostructures.
  In fact, this is most likely the reason why in a previous paper on 2DEG at 
ferroelectric interfaces only one polarization orientation was found to be 
stable.\cite{28}
   
   At this point, it should be noted that according to the model,\cite{29} 
the critical thickness for the stabilization of ferroelectricity is proportional 
to the band gap $\Delta$, which is known to be severely underestimated by 
the local density approximation of the exchange-correlation functional.
  Thus, we expect that the actual transition would take place at thicknesses 
of the order of $\sim 30$ unit cells of PTO, which is still in the range of typical 
values grown experimentally.
   
\subsection{Free carriers}

\begin{figure*}
\begin{center}
\includegraphics[width=2\columnwidth]{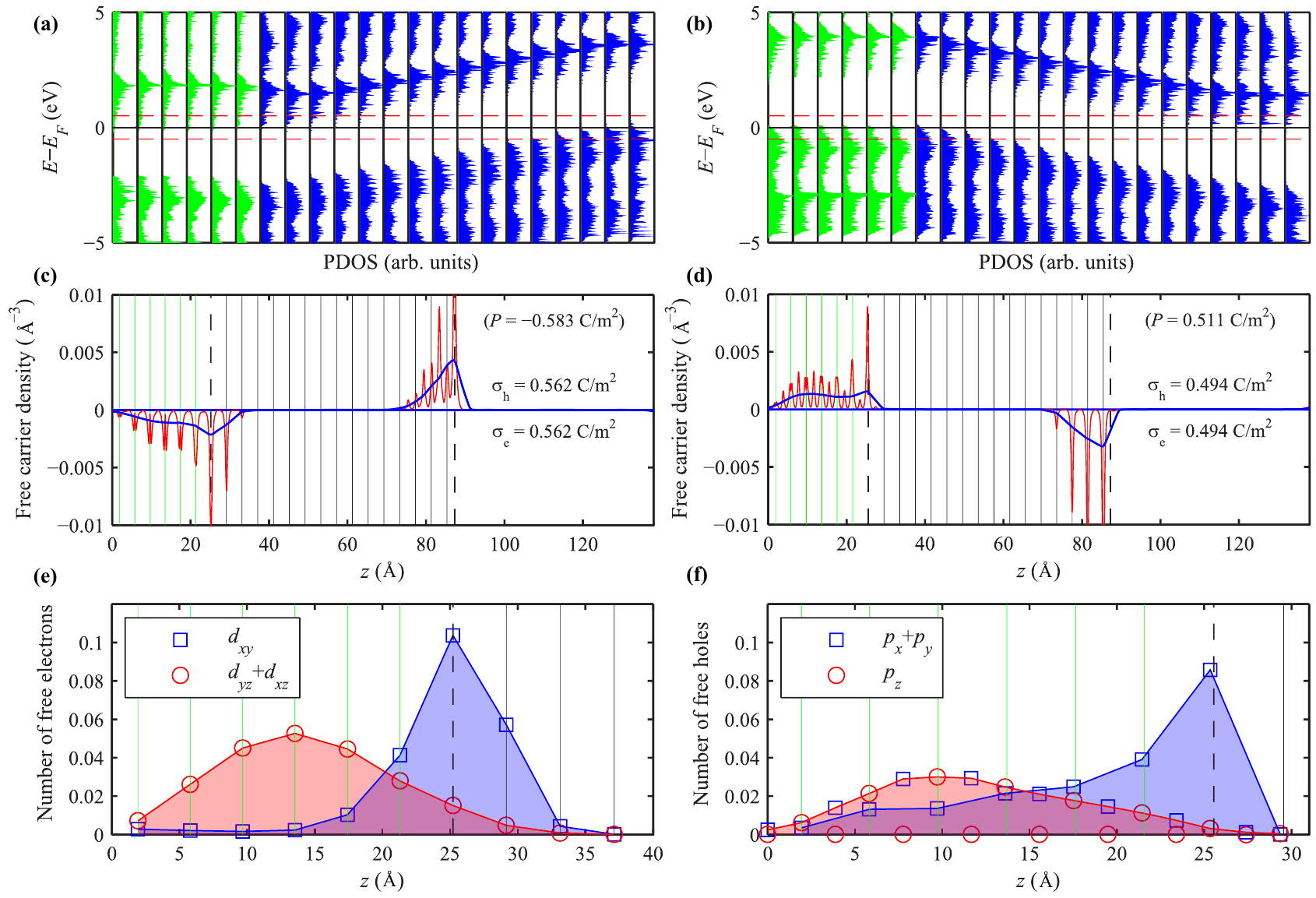}
\caption{\label{fig4} 
  Electronic structure of the PTO/STO interface with $m = 16$ with both 
downward (left panels) and upward (right panels) polarizations.
  The bilayer-resolved PDOS is plotted in (a) and (b), with green and blue 
representing STO and PTO, respectively. 
  $E_F$ is the Fermi energy of each structure.
  The red dashed lines indicate the energy window used in the free 
carrier density mapping (see Appendix C).
  In (c) and (d), plane-averaged (red) and macroscopically averaged\cite{46} 
(blue) profiles of free charge are plotted, with the positive and negative values 
representing the free holes and free electrons, respectively.
  Free carrier densities in STO are further resolved into orbital populations 
as shown in (e) and (f). In (c)-(f), the positions of interface and surface are 
indicated by the thick dashed lines, while thin solid lines mark the position 
of every Ti atom, with green and black for those inside STO and PTO, 
respectively. } 
\end{center}
\end{figure*}
   
  In Figs. 4(a) and 4(b) we plot the projected density of states (PDOS) of 
each unit-cell bilayer for the $m = 16$ interface and both polarization 
orientations.
  These figures confirm the occurrence of an electronic reconstruction, 
with the conduction (valence) band at the interface and the valence (conduction) 
band at the surface crossing the Fermi level of the heterostructure with downward (upward) polarization. 
  Additionally, the remnant depolarization field resulting from the incomplete 
screening, responsible for the tilting of the bands, can also be clearly seen 
from the PDOS.
   
  Furthermore, the distribution of free-charge carriers can be mapped using 
the local density of states (technical details can be found in the Appendix C).
  In Figs. 4(c) and 4(d), we plot the plane-averaged and macro-averaged\cite{46} 
free-carrier density for both $P_{\downarrow}$ and $P_{\uparrow}$.
  The two profiles display similar features (with opposite sign).
  The sheet of free-charge at the surface is strongly confined in both cases, 
as a consequence of the remnant depolarizing field in the ferroelectric layer.
  On the STO side, charge is more loosely bound to the interface by the electric 
field in the PTO layer and its own electrostatic interaction.\cite{47}
  Therefore, it peaks near the interface and decays slowly towards the 
bottom surface.
  One detail that is worth noting is that while the 2DEG in the prototypical 
LAO/STO system is strictly confined to the STO side due to the band alignment 
between the two materials,\cite{48} here the almost vanishing band offset at 
the interface means that some free charge spreads into the first layers of PTO. 
   
   The free-charge profiles plotted in Figs. 4(c) and 4(d) can be integrated 
along $z$ to compute the free-carrier density per unit area, labeled as
$\sigma_h$ and $\sigma_e$ for holes and electrons, respectively.
  The magnitudes of $\sigma_h$ and $\sigma_e$ match perfectly for 
both polarization directions, as displayed in Figs. 4(c) and 4(d), obeying 
charge neutrality.
  Also, they are very close to the value of the stable polarization inside 
the PTO layer, in agreement with the prediction of the model presented 
in Ref.~\onlinecite{29}.
  The almost perfect overlap between polarization and interfacial 
  free -charge values results in an excellent screening of the depolarization effects, 
enabling the stability of the ferroelectric phase.
  This result is very robust, and it displays very little sensitivity to the atomic 
relaxations of the STO (hosting the 2-dimensional free-charge gas at the 
interface) or the confinement of the free charge, as discussed in Appendix A.
   
   Moreover, in addition to the agreement between the integral values 
of the free charges ($\sigma_h$ and $\sigma_e$) and the polarization ($P$), 
their spatial distributions are found to match as well.
  This can be observed either by differentiating the polarization to obtain 
the bound charge profile and comparing it with the free-charge distribution,\cite{42} 
or, alternatively, integrating the free-charge density to obtain the electric 
displacement profile and comparing it with the polarization one.
  Taking a point in the vacuum region as one of the limits in the domain of 
integration (since in the vacuum region the electric displacement $D = 0$, 
as guaranteed by the dipole correction), one obtains the electric displacement 
profiles represented as black dashed curves in Fig. 3(a).
  Since the difference between the polarization and the electric displacement 
is the electric field, the excellent overlap between the two curves throughout 
the whole heterostructure reflects the excellent screening provided by the 
electronic reconstruction.
  The remnant electric field inside the PTO layer can be estimated in a rather 
indirect way using the expression $(P-\sigma)/\epsilon_0$ and the values 
listed in Figs. 4(c) and 4(d).
  For both polarization orientations, this calculation yields a remnant electric 
field of $\sim 0.2$ V/\AA.
  This value is susceptible to a large error since it is obtained from the 
subtraction of two very large and similar quantities.
  A direct estimation of the remnant depolarization field can be obtained 
from the macroscopic average\cite{46} of the electrostatic potential, 
shown in Fig. 5, resulting in a value for the remnant electric field of 
$\sim 0.05$ V/\AA.
  The consistency of the electrostatic analysis constitutes strong evidence 
that the driving force for the electronic reconstruction is indeed the polar 
discontinuity, in excellent analogy with the case of the polar LAO/STO interface.
   
\begin{figure}
\begin{center}
\includegraphics[width=\columnwidth]{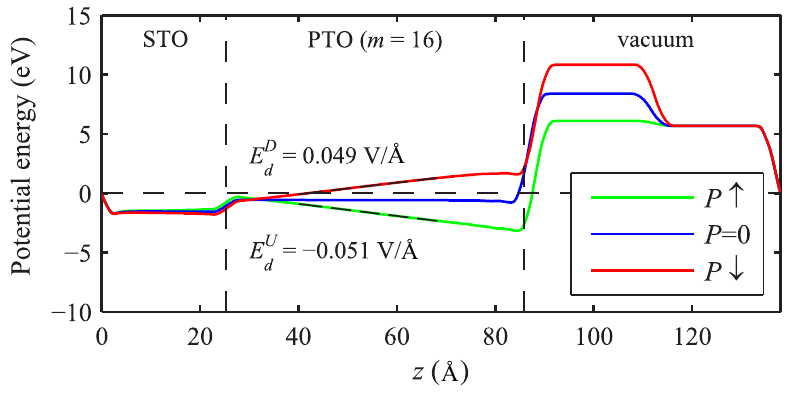}
\caption{\label{fig5} 
  Macroscopic average\cite{46} of the electrostatic potential energy for each 
polarization state in $m = 16$ heterostructures.
  The black dashed lines that overlap with the red and green curves indicate 
the linear fitting, which yields the values of the remnant depolarization field,  
and   for $P_{\downarrow}$and $P_{\uparrow}$, respectively.} 
\end{center}
\end{figure}
   
\subsection{2DHG in SrTiO$_3$}
 
 \begin{figure}
\begin{center}
\includegraphics[width=\columnwidth]{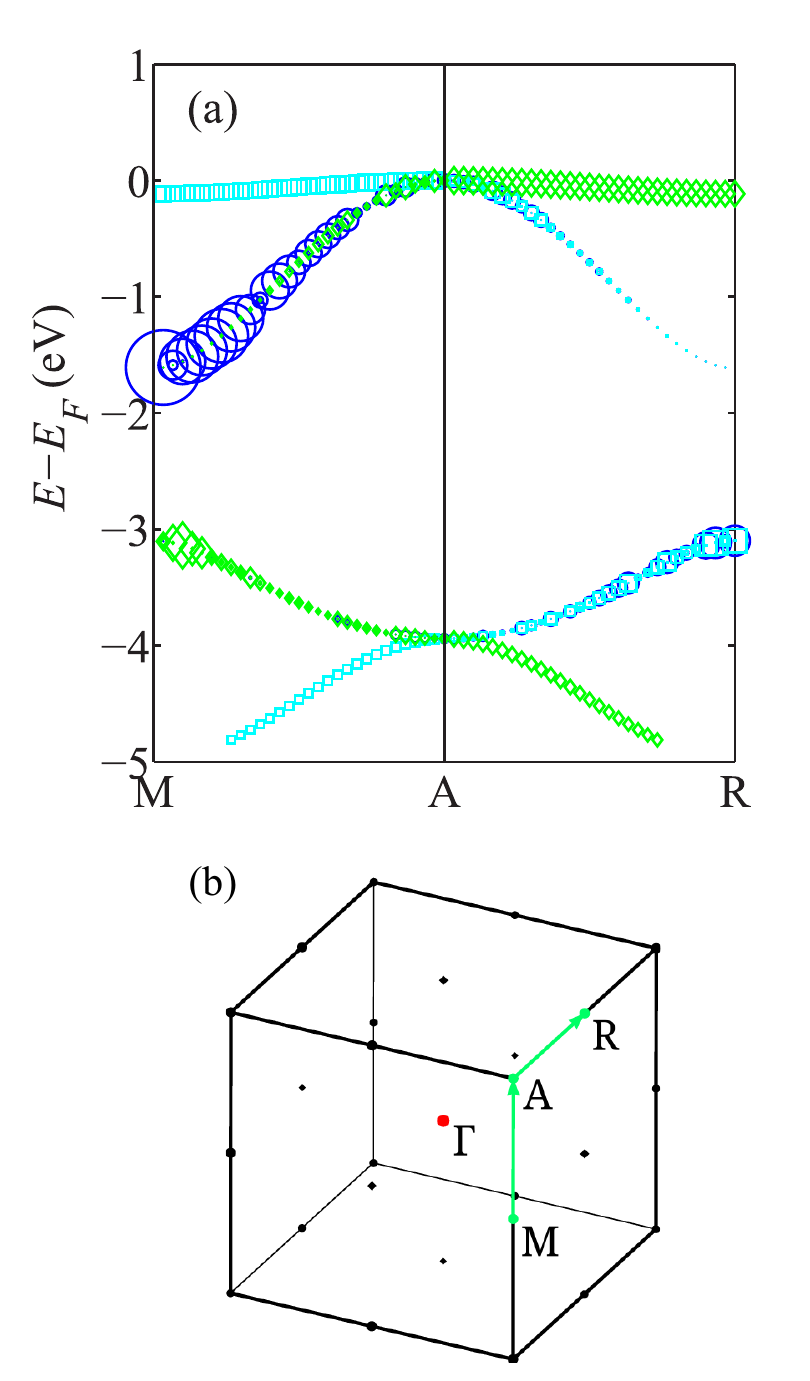}
\caption{\label{fig6} 
(a) Dispersion of the valence bands in STO decomposed into contributions 
from the different O 2$p$ orbitals: $p_x$ of SrO planes in blue circles, $p_x$ of 
TiO$_2$ planes in cyan squares, and $p_z$ of TiO$_2$ planes in green
 diamonds.
  The size of the symbols represents the weight of each orbital in the 
corresponding eigenstate.
  $E_F$ is the Fermi energy of bulk STO. 
  In (b), the $k$-point path in the tetragonal Brillouin zone is depicted. 
  Even if STO is cubic in the bulk, here we chose to plot the tetragonal 
representation to illustrate the interface-induced symmetry breaking 
between $z$ and in-plane ($x$ or $y$) directions.} 
\end{center}
\end{figure}
 
  It is interesting to discuss in more detail the electronic structure of the 
interface and, in particular, the differences between 2DEG and 2DHG, 
since their distinct properties might constitute a route for the design of devices 
in which the orientation of the polarization could be used to manipulate the 
transport characteristics in a non-volatile fashion.
  For $P_{\downarrow}$, a large amount of free electrons accumulates at 
the conduction band of STO, which means that both $d_{xy}$ and 
$d_{xz}+d_{yz}$ of the Ti atoms are being partially occupied.
  As it was demonstrated by Popovic {\it et al.},\cite{49} these two sets 
of orbitals spread very differently from the interface in LAO/STO, an 
effect that, they propose, should importantly affect electronic transport 
in the 2DEG.
  The same behavior is observed here, as evidenced in Fig. 4(e).
 
 \begin{figure*}
\begin{center}
\includegraphics[width=2\columnwidth]{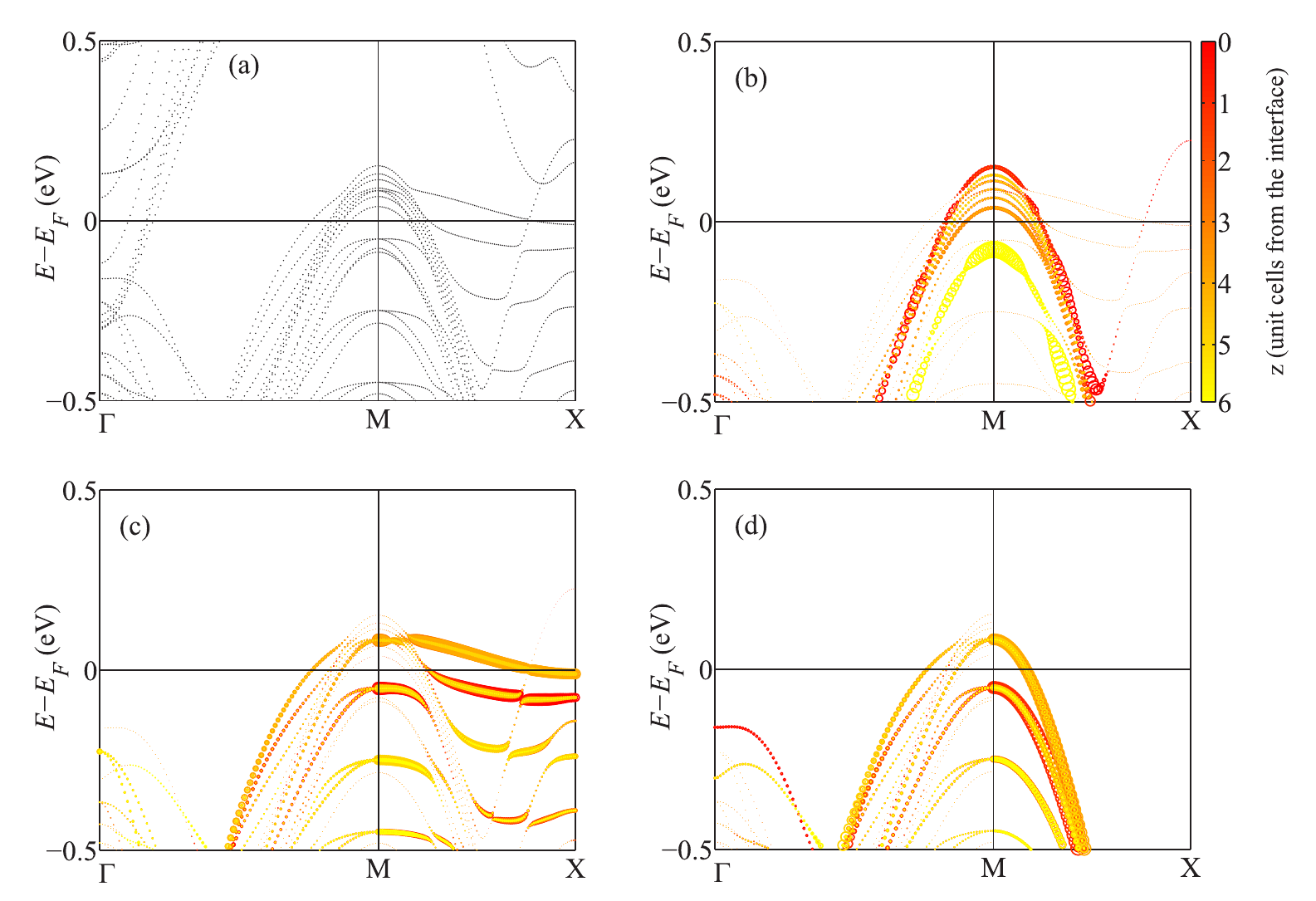}
\caption{\label{fig7} 
Band structure of the slab with $m = 16$ and upward polarization. 
  (a) Electronic bands around the Fermi energy, where the top of the valence 
band of STO is seen crossing the Fermi level at M in the 2D Brillouin zone. 
  Here, the $k$-point path from M to X corresponds to the projection of the 
path A to R in Fig. 6(b) onto the 2D Brillouin zone.
  The valence bands of STO are decomposed into contributions from different 
orbitals, with (b), (c) and (d) corresponding to the projection over non-bonding 
$p_x$ and $p_y$ orbitals in TiO$_2$ layers, $p_x + p_y$ in SrO layers, and 
$p_z$ in TiO$_2$ layers, respectively.
  In (b)-(d), each eigenvalue $E_n(\mathbf{k})$ is plotted as a circle with the 
size representing the weight of each oxygen $p$ orbital, and the color indicating 
the distance from the interface, as illustrated by the legend in (b). 
  $E_F$ is the Fermi energy of the heterostructure.} 
\end{center}
\end{figure*}
 
  Interestingly, here, the reversible polarization allows a 
similar analysis to be carried out for 2DHG.
  Even though a 2DHG should also form at the $p$-interface of a pristine 
LAO/STO system, the fact that it has never been observed experimentally 
has discouraged a deeper theoretical analysis of this system.
  However, since ferroelectric thin films are typically grown at temperatures 
higher than the Curie one, the electric-field-induced donor-acceptor defect 
pairs that are believed to be responsible for the insulating nature of the 
$p$-interface\cite{23} are probably energetically unfavorable during the 
deposition of the ferroelectrics.
  Therefore, we propose that if a $P_{\uparrow}$ monodomain can be 
condensed after growth, intrinsic electronic reconstruction or surface 
redox reactions are more likely to provide the necessary screening charge 
during the cooling process and consequently the 2DHG might survive in 
these systems, as opposed to what happens in LAO/STO.
  Here, we take advantage of the reversible polarization to analyze in 
more detail the electronic structure of the interface in the presence of a 2DHG.
  
  A decomposition of the free charge into different orbital components 
reveals a behavior similar to the one described in Refs.~\onlinecite{49} and 
\onlinecite{50} for 2DEG, in which O 2$p$ orbitals form two sets of bands 
that behave similarly to the split $t_{2g}$ orbitals in the 2DEG.
  By examining every individual O atom in STO, we find that all 2$p$ 
orbitals along Ti-O bonds are fully occupied (no holes).
  The same feature has just been reported recently for doped STO in 
the bulk.\cite{51}
  Then we decompose the PDOS corresponding to the 2DHG into the 
$p_z$ and $p_x+p_y$ contributions in Fig. 4(f), where the points can 
be more easily interpreted by classifying them into three groups: 
$(i)$ the population of the bands with $p_z$ character from the SrO 
planes is zero, since they are along the Ti-O bonds; 
$(ii)$ the bands with character $p_z$ from TiO$_2$ and $p_x+p_y$ 
from SrO behave similarly, with both populations peaking inside STO 
(red shadow), just like the free electrons in the $d_{xz}+d_{yz}$ orbitals in 
Fig. 4(e); 
and $(iii)$ the population of the $p_x+p_y$ bands in TiO$_2$ atomic 
planes peaks at the interface (blue shadow), in correspondence with
$d_{xy}$ in Fig. 4(e) (note that in this case for every O atom either 
$p_x$ or $p_y$ is full since it lies along the Ti-O bond).
  According to this, the orbital distribution in 2DHG that forms in the 
$P_{\uparrow}$ case bears a strong resemblance with the one of  
2DEG, i.e., a very localized band right at the interface (Ti $d_{xy}$ for 
 2DEG, TiO$_2$ $p_x+p_y$ for 2DHG) and a long tail formed 
by bands with a different character (Ti $d_{xz}+d_{yz}$ for 2DEG, 
TiO$_2$ $p_z$ and SrO $p_x+p_y$ for 2DHG).
  
  These observations can be used to formulate a set of rules that 
determine the free charge distribution in these systems, and reveal the 
origin of the common features shared by 2DEG and 2DHG.
  As it was already discussed in Refs.~\onlinecite{49} and \onlinecite{50}, 
the splitting of the $t_{2g}$ levels of Ti atoms near the interface leads to 
a clear differentiation between the electronic population of two sets of 
bands.
  In those papers the analysis focused on the conduction band of STO, 
and it was found that the $d_{xy}$ orbitals form bands with very small 
hopping along $z$ and, therefore, a strong 2D localization.
  The other set of bands, which contributes to the charge spreading along 
$z$, is formed by the linear combination of $d_{xz}$ and $d_{yz}$ orbitals 
from different Ti layers.
  Interestingly, this behavior can also be inferred from the band structure 
of bulk STO.
  In the bulk, even if degenerated at the bottom of the conduction band 
(located at $\Gamma$), $d_{xz}$ and $d_{yz}$ on one hand and $d_{xy}$ 
on the other form two separate bands when looking at the dispersion along $z$.
  The negligible dispersion of the $d_{xy}$ band along $z$ hints at the lack 
of mixing between different Ti layers at the interface.
  This strong 2D localization implies a strong sensitivity to the local 
electrostatic potential and a shift of the bands with $z$, as observed in 
Refs.~\onlinecite{49} and \onlinecite{50}.
  Instead, the delocalization of the $d_{xz}+d_{yz}$ bands gives rise to 
extended bands at the interface, which are less sensitive to the potential well that 
confines the 2DEG.

  A parallel argument can be made for the charge distribution in 2DHG.
  As shown in Fig. 6(a), the bands at the top of the valence band can be 
classified into two categories: a band localized along $z$ formed by 
non-bonding oxygen $p_x+p_y$ orbitals at TiO$_2$ planes, and another 
one, dispersive along $z$, formed by $p_x+p_y$ at SrO planes and $p_z$ 
at TiO$_2$ ones.
  This decomposition coincides with the different populations seen in the 
analysis of the free-charge profiles, and it is in agreement with the interpretation 
given for the population distribution in the 2DEG.
  The definite confirmation that the argument about band localization explains 
the population distribution is found by observing the electronic structure of the
interface.
  In Fig. 7, the electronic bands near the Fermi level are decomposed into 
contributions from different oxygen $p$ orbitals (each one in a separate panel) 
and atomic layers (the color gradient goes from red for the TiO$_2$ plane at 
the interface to yellow for the SrO surface).
  This decomposition shows that there is a set of bands formed by $p_x$ 
orbitals at individual TiO$_2$ planes that shifts considerably with $z$ [Fig. 7(b)]. 
  Several of these bands cross the Fermi level, but their population should 
decay rapidly from the interface.
 Then, there is essentially one band, with contributions from the rest of the
 non-bonding $p$ orbitals throughout the whole STO substrate [Fig. 7(c) 
 and 7(d)], responsible for the smoother charge profile in Fig. 4(f). 
  
  The analysis of the different charge populations just described allows us to
anticipate, at least qualitatively, the electronic structure of 2DEG or 2DHG 
that might form at other polar interfaces from the bulk band structure of the 
host materials.
  It can be used, for instance, to estimate the effective masses of free-charge 
carriers in the 2DHG from calculations of the bulk band structure.
  From the in-plane dispersion of the $z$-localized band we get an effective 
mass of $m_x^* = m_y^* = -1.17 m_e$ if calculated from bulk [cyan squares 
in Fig. 6(a)] and $-0.94 m_e$ if calculated directly from the interface band 
structure.
  This constitutes a reasonable agreement, especially considering the difficulty 
involved in isolating the right band from the interface band structure.
  In bulk, one of the $z$-extended bands (blue circles) is degenerated with the 
$z$-localized one (cyan squares) in one of the in-plane directions, while the 
other one (green diamonds) has a much larger effective mass, as shown in 
Fig. 6(a).
  For the heavy holes of the $z$-extended band, the bulk and interface band 
structure calculations yield values of $-13 m_e$ and $-12.7 m_e$, respectively.
  
  This analysis was already applied to obtain the density of states
  used for all the calculations in Ref.~\onlinecite{29}
  from the band structure of bulk PTO.
  The validity of the model, and the parameters used for its predictions, was 
confirmed after the comparison with the test-case first-principles simulation 
of PTO free-standing slabs.\cite{29}

\section{Discussion}

  The first-principles study presented here only explores the competition 
between the monodomain phase sustained by an electronic reconstruction 
and a paraelectric configuration.
  However, in realistic experiments, additional elements might come into play.
  First, it is widely acknowledge that the source of the free charge accumulated 
at polar interfaces is at least partially provided by redox processes at either 
the surface, the interface, or a combination of both.\cite{22,23}
  The participation of other sources of free charge could modify the balance 
between different phases, but it should not affect the main conclusion 
of this paper, i.e. the possibility of stabilizing monodomain ferroelectricity 
by two-dimensional free-carrier gases.
  The only difference is that in that case, the effective gap $\Delta$ will 
encompass things like the formation energy of defects and the defect-defect 
interaction.\cite{29}
  This means that for both electronic reconstruction and surface 
electrochemical processes, the phenomenology is expected to be similar 
and the main quantitative difference would be the critical thickness for the 
stable ferroelectric phase.
  In fact, as mentioned above, since ferroelectric thin films are usually grown 
above the Curie temperature, during the deposition process there is no 
polarization discontinuity at the interface and therefore no electric-field-i
nduced redox reactions.
  This should help interfaces and surfaces at ferroelectric thin films to be 
cleaner than those in the case of LAO,\cite{18} for which the polarization 
mismatch is present at all times.
  In the case of ferroelectrics, most mechanisms providing free charge 
should take place during the cooling of the sample, when the 
heterostructure is fully formed, and this probably favors intrinsic 
processes over extrinsic ones. 
  
  Secondly, the formation of polarization polydomains is an intrinsic 
screening mechanism that is always accessible in these systems and 
that eliminates the necessity for free-charge accumulation at the interfaces.
  The competition between these two screening mechanisms was analyzed 
in Ref.~\onlinecite{29}, demonstrating a crossover with thickness between 
the two configurations: polydomain for small thicknesses and monodomain 
with 2DEG for larger thicknesses.
  More importantly, the stability of monodomain ferroelectricity in PTO thin 
films on STO has indeed been confirmed experimentally numerous 
times,\cite{30,31,32,33,34,35} but in the absence of screening charge at 
the interface, such a configuration would be impossible according to simple 
electrostatic arguments.
  Surprisingly, the process that stabilizes monodomain ferroelectricity in 
such systems has received little attention in the past.\cite{52}
  This paper provides a possible mechanism that would explain such 
observations, and it should motivate further investigation of the screening 
processes that take place in these systems and how one might take 
advantage of them to confer ferroelectric interfaces with new functionalities.
  
  Finally, one important result of this paper is that it provides strong evidence 
supporting the polarization mismatch as the driving force for the formation 
of the 2DEG at the LAO/STO interface.
  This mechanism is still under debate, mostly because of the predictions 
from alternative interpretations that have been proposed, like intrinsic
doping introduced by the LaO$^+$ layer at the interface,\cite{53} 
which coincide with those obtained with the polar catastrophe model.
  However, such interpretations would fail to explain the results of this 
paper, where the origin is clearly the polar catastrophe.

\section{Conclusions}
  
  In this paper, we have systematically studied the properties of 2DEG at a 
prototypical ferroelectric interface, PbTiO$_3$/SrTiO$_3$, finding that, 
above a critical thickness, the ferroelectric monodomain can be stably 
sustained by the screening of the depolarization field provided by either 
a 2DEG or a 2DHG at the interface, depending on the polarization orientation.
  In this regime, the system possesses a tri-stable energy landscape in 
which two polar and metallic states, and one non-polar and insulating state 
are accessible.
  All these results agree very nicely with the predictions of the model 
described in Ref.~\onlinecite{29}, including the discontinuous switching
of ferroelectricity with thickness.
  The analysis of the electronic structure of the system has revealed 
some interesting common features between the electron and hole 
population distributions in the 2DEG and 2DHG, respectively, allowing us
to outline some simple rules to predict basic properties of these systems 
from bulk characteristics of the constituent materials.
  
  The formation of a 2DEG at this very well-known heterostructure may 
open the door to the design of structurally new, relatively simple, all-oxide, 
field-effect non-volatile devices thanks to the retention provided by the 
spontaneous polarization.
  The fact that the process is driven by electrostatics means that it is not 
material-specific, and this idea can be exploited to engineer new functional 
interfaces or enhance the performance of the device by combining ferroelectric 
or multiferroic films with, for instance, magnetic substrates.

\begin{acknowledgments}  
  Computations were performed at the National Supercomputer Center in 
Tianjin (NSCC-TJ), the Spanish Supercomputer Network (RES) and 
computational resources at the Donostia International Physics Center.
  This paper was supported by the National Natural Science Foundation of 
China (Grants No. 11222218 and No. 11321202), Natural 
Science Foundation of Zhejiang Province (Grant No. LZ14A020001), the Fundamental 
Research Funds for the Central Universities, and Spain's 
Ministry of Economy and Finance (MINECO; Grant No. 
FIS2012-37549-C05).
  Atomic configurations were visualized by VESTA.\cite{54}
\end{acknowledgments}

\appendix

\section{Effect of atomic relaxations on the amount of free charge 
and its profile}

\begin{figure}
\begin{center}
\includegraphics[width=\columnwidth]{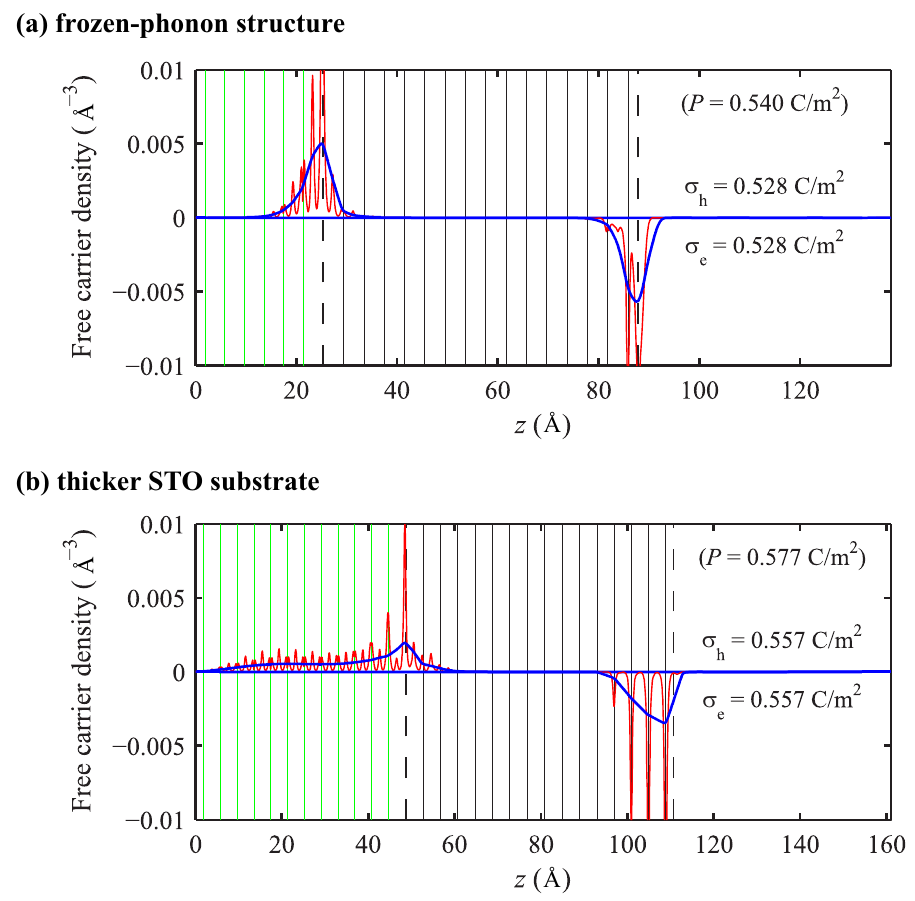}
\caption{\label{fig8} 
  Plane-averaged (red) and macroscopically averaged (blue) profiles of 
free charge for $m =16$ interfaces. 
  (a) The frozen-phonon calculation at the local energy 
minimum of Fig. 2, with upward polarization. 
  (b) The thickness of the STO substrate is 12 unit cells, instead of 6 
as in the rest of the article.
  The positions of interface and surface are indicated by the thick dashed 
lines, while thin solid lines mark the position of every Ti atom, with green 
and black for those inside STO and PTO, respectively.} 
\end{center}
\end{figure}

  Among the systems studied here, those in which the ferroelectric material
possesses a finite polarization have metallic interfaces, regardless of 
whether they are relaxed or generated using the frozen-phonon scheme.
  In addition, also in all of them, the amount of free charge accumulated 
at the interface compensates almost completely for the polarity of the interface.
  This is the consequence of the huge energy penalty that an unscreened 
depolarizing field constitutes.
  The reduction in the electrostatic energy easily compensates for the cost of 
transferring charge from the valence to the conduction band and, therefore, 
for typical materials, free charge accumulates until its value almost reaches 
that of the polarization.

  As discussed in the paper, this phenomenon is driven by electrostatics 
and some of the fundamental fingerprints, like the critical thickness or the 
magnitude of the polarization and free charge, depend mostly on the bulk 
properties of the ferroelectric.
  This is the main reason for the agreement between the relaxed structures 
and the frozen-phonon analysis.
  The relaxations of the STO layer play a secondary role in the stabilization 
of the ferroelectric phase.
  In general, in the model described by Eq. (3), the influence of the substrate 
is restricted to the density of states, $g$, and (possibly) the effective band gap, 
$\Delta$.
  Even though, in general, the structural relaxations near the interface can 
affect the value of $\Delta$, causing disagreements between estimations 
derived from a frozen-phonon analysis and the fully relaxed structures, 
for this particular system, we find that neither the effective band gap of the 
system nor the DOS at the interface is particularly sensitive to the 
atomic relaxations, guaranteeing the compatibility of the two methods. 
  Figure 3(b) demonstrates the agreement in the critical thickness and 
the evolution of the polarization with the thickness of the ferroelectric material.
  The agreement is also notable in the free charge accumulated at the 
interface, which amounts to 0.49 C/m$^2$ [Fig. 4(d)] in the relaxed 
structure and 0.53 C/m$^2$ [Fig. 8(a)] in the frozen-phonon one.
  
  On the other hand, the ionic relaxations of the STO layers play a 
 mayor role in the width of the resulting two-dimensional free-carrier 
gases.
  It is evident from the comparison of Fig. 4(d) and Fig. 8(a) that the 
reduced susceptibility of STO in the frozen-phonon calculations (only 
the electrons contribute to the material polarizability) results in a stronger 
confinement of the free charge.
  In fact, due to the very large susceptibility of STO, the 2DEG or the 
2DHG tends to penetrate very deep inside the substrate;\cite{47} 
  this can be appreciated in Fig. 8(b). 
  Additionally, Fig. 8(b) also shows that the values of the stable polarization 
and free charge, as well as the qualitative aspects of the 2DHG distribution, 
are well converged for the thickness of 6 unit cells of STO used throughout 
the paper. 

\section{Polarization fluctuations in ab initio fully relaxed structures}

  Due to the shallowness of the energy minimum corresponding to the 
ferroelectric phase for thicknesses near the transition, any finite value of 
the force threshold in the atomic relaxations entails some fluctuations in 
the atomic coordinates and thus in the values of the polarization of the system.
  The phenomenological model can be used to obtain an estimation of the 
magnitude of these fluctuations, linking changes in polarization to a threshold  
in the forces.
  Assuming that the force tolerance is the same for any atom for a displacement 
given by a small amplitude of the soft-mode distortion, then a change in energy 
can be expressed as a line derivative with respect to the amplitude of the 
soft-mode deformation. 
  Using this premise, the following relation can be found
\begin{align}
\delta P &= {\partial P \over \partial P_{\eta} } \delta P_{\eta} \\ &= 
{\partial P \over \partial P_{\eta} } (-mc) \left ( e \sum_{i=1}^5 Z_i \xi_i
{\partial^2 G \over \partial P_{\eta}^2} \right )^{-1} 
\delta \left ( \sum_{i=1}^5 F_i \xi_i \right ) \; . \nonumber
\end{align}  
Here, $F_i$ and $\xi_i$ are the force and soft-mode displacement of 
the $i$-th atom in a perovskite unit cell, $P_{\eta}$ is the soft mode 
contribution to the polarization, $G$ is the free energy per unit area, 
$c$ is the out-of-plane lattice constant of bulk PTO, $m$ is the thickness 
of the PTO thin film, and $Z_i$ is the Born effective charge associated 
to the $i$-th atom.
  Using $\delta F_i = 0.025$ eV/\AA\ as the force threshold used in the 
ab initio optimizations, and using the curvature of the energy curves at the 
equilibrium ferroelectric configuration, $\partial^2 G / \partial P^2$, from 
the fitting of the frozen phonon calculations, we get the values plotted 
as error bars in Fig. 3.

\section{Mapping of free carrier density} 

  The spatial distribution of the free carriers can be mapped in a 
quantitative way by extending the method used in Ref.~\onlinecite{42}.
  First, the local density of states, $\tilde{\rho} (\mathbf{r}, E)$, which is 
a function of the spatial coordinates {\bf r} and the energy $E$, is computed 
using the same occupation function (the Fermi-Dirac distribution, $f_{\rm FD}$, 
for electrons and $1- f_{\rm FD}$ for the holes) and smearing temperature 
as during the self-consistent computation of the one-particle eigenstates.
  Then, the local density is integrated over an energy window, as indicated 
by the red dashed lines in Figs. 4(a) and 4(b).
  It should be noted that by integrating   for the electrons in the energy 
window depicted in Fig. 4(a), one obtains contributions from the conduction 
band near the interface as well as from the valence band near the PTO surface.
  With an appropriate choice for the energy window, these two contributions 
are spatially separated while all partially occupied states are included.
  Then the contribution to the electrons coming from the valence band can 
be removed, and only the part corresponding to the free electrons is plotted, 
as shown in Fig. 4(c).
  The same method is used to plot the free holes in Fig. 4(c) and free electrons 
and holes in Fig. 4(d).
  In this paper, we show the free-carrier densities that are calculated with an 
energy window spanning from -0.5 eV to 0.5 eV in Figs. 4(c) and 4(d), 
while integration from -0.6 eV to 0.6 eV gives exactly the same results.
  This indicates that our choice of the energy window meets the prerequisites 
mentioned above and should yield meaningful results of the free carrier density.

\begin{figure}
\begin{center}
\caption{\label{XXX} Caption.} 
\end{center}
\end{figure}

\end{document}